\documentclass[12pt]{article}
\usepackage{latexsym,epsfig,amssymb,amsmath}
\usepackage{color}
\usepackage[all]{xy}
\newcommand{\ft}[2]{{\textstyle\frac{#1}{#2}}}

\def\bfone{\relax{\rm 1\kern-.35em 1}}

\newcommand{\be}{\begin{equation}}
\newcommand{\ee}{\end{equation}}
\newcommand{\ben}{\begin{displaymath}}
\newcommand{\een}{\end{displaymath}}
\newcommand{\bea}{\begin{eqnarray}}
\newcommand{\eea}{\end{eqnarray}}
\newcommand{\nn}{\nonumber}

\newcommand{\bean}{\begin{eqnarray*}}
\newcommand{\eean}{\end{eqnarray*}}
\newcommand{\beqs}{\begin{eqnarray}}
\newcommand{\eeqs}{\end{eqnarray}}

\newcommand{\e}{\epsilon}
\newcommand{\eb}{\overline{\epsilon}}
\newcommand{\es}{\epsilon^\star}
\newcommand{\esb}{\overline{\epsilon^\star}}
\newcommand{\s}{\star}

\newcommand{\mathon}{\mathversion{bold}}
\newcommand{\mathoff}{\mathversion{normal}}

\makeatletter
\@addtoreset{equation}{section}
\makeatother

\newcommand{\w}[1]{\\[0.#1cm]}
\newcommand{\eq}[1]{(\ref{#1})}

\def\ft#1#2{{\textstyle{\frac{\scriptstyle #1}{\scriptstyle #2} } }}

\begin{document}

\thispagestyle{empty}

\begin{flushright}\small
MIFPA-13-32
\end{flushright}

%%%%%%%%%%%%%%%%%%%%%%%%%%%%%%%%%%%%%%%%%

\bigskip
\bigskip

\mathon
\vskip 10mm

\begin{center}

  {\Large{\bf Supersymmetric Warped AdS in \\[1.8ex]    Extended Topologically Massive Supergravity}}

\end{center}

\mathoff

%%%%%%%%%%%%%%%%%%%%%%%%%%%%%%%%%%%%%%%%

\vskip 6mm

\begin{center}

{\bf N.S. Deger$^a$, A. Kaya$^b$, H. Samtleben$^c$ and E. Sezgin$^d$}\\
\vskip 4mm

$^a$\,{\em Bogazici University, Department of
Mathematics, \\ 34342, Bebek, Istanbul, Turkey} \\
\vskip 4mm

$^b$\,{\em Bogazici University, Department of
Physics, \\ 34342, Bebek, Istanbul, Turkey} \\
\vskip 4mm

$^c$\,{\em Universit\'e de Lyon, Laboratoire de Physique, UMR 5672, CNRS et ENS de Lyon,\\
46 all\'ee d'Italie, F-69364 Lyon CEDEX 07, France} \\
\vskip 4mm

$^d$\,{\em George P. and Cynthia W. Mitchell Institute \\for Fundamental
Physics and Astronomy \\
Texas A\&M University, College Station, TX 77843-4242, USA}\\

\end{center}

\vskip1.2cm

\begin{center} {\bf Abstract } \end{center}

\begin{quotation}\noindent

We determine the most general form of off-shell $N=(1,1)$ supergravity field configurations in three dimensions by requiring that at least one off-shell Killing spinor exists. We then impose the field equations of the topologically massive off-shell supergravity and find a class of solutions whose properties crucially depend on the norm of the auxiliary vector field. These are spacelike-squashed and timelike-stretched $AdS_3$ for the spacelike and timelike norms, respectively.
At the transition point where the norm vanishes, the solution is null warped $AdS_3$. This occurs when the coefficient of the Lorentz-Chern-Simons
term is related to the $AdS$ radius by $\mu\ell=2$.
We find that the spacelike-squashed $AdS_3$ can be modded out by  a suitable discrete subgroup of the isometry group, yielding an extremal black hole solution which avoids closed timelike curves.

\end{quotation}

\newpage
\setcounter{page}{1}

%\tableofcontents

\bigskip

%%%%%%%%%%%%%%%%%%%%%%%%%%%%%%%%%%%%%%%%%
\section{Introduction}
%%%%%%%%%%%%%%%%%%%%%%%%%%%%%%%%%%%%%%%%%

It is well known that the low energy limit of string theory can be described by supergravity theories and higher order curvature terms appear in a derivative expansion in the slope parameter $\alpha'$. Only the two derivative Lagrangian is exactly supersymmetric and beyond that supersymmetry holds only order by order in $\alpha'$. The higher derivative terms, including the curvature squared terms, need to be treated as interactions that become relevant at increasingly higher energies. On the other hand, off-shell supergravities exist in lower dimensions as long as the number of supersymmetries does not exceed eight. In this case, given that the local supersymmetry transformations do not depend on field equations, one can construct higher derivative invariants that are exactly supersymmetric by themselves. Another key property of these theories is that the auxiliary fields, characterized by having algebraic field equations in a 2-derivative off-shell supergravity, typically become propagating fields once
higher derivative invariants are added.

The properties of the off-shell higher derivative supergravities stated above raise the interesting possibility of treating them in their own right, postponing the questions of their relation to on-shell supergravities that arise in string theory in a suitable background. Once this philosophy is adopted, it is natural to look for exact solutions of these theories, starting with the maximally symmetric ones. In particular, seeking supersymmetric solutions is aided greatly by studying the off-shell Killing spinor equation.

The simplest off-shell supergravity that contains an auxiliary field is the off-shell $N=(1,0)$ supergravity in three dimensions \cite{Uematsu:1984zy}. In addition to the graviton and a single Majorana gravitino, it contains an auxiliary real scalar field. The analysis of the Killing spinor equations of
on-shell topologically massive supergravity \cite{Deser:1982sv,Deser:1982sw}
was given in \cite{Gibbons:2008vi}, and the off-shell Killing spinor equations were analyzed in \cite{Andringa:2009yc}, where the most general form of the metric and auxiliary field consistent with supersymmetry was determined. Using these results in field equations that follow from particular $N=(1,0)$ off-shell higher derivative supergravity models, exact solutions for which the auxiliary field is constant were found and examined in detail in \cite{Andringa:2009yc,Bergshoeff:2010mf}. Supersymmetric solutions of these models are necessarily of the pp-wave type. Supersymmetric warped $AdS_3$ solutions on the other hand have appeared so far only in on-shell extended $N=(2,0)$ supergravity coupled to a vector multiplet~\cite{Banados:2007sq}.

In this paper, we shall study solutions of the Killing spinor equations associated with $N=(1,1)$ off-shell supergravity in three dimensions \cite{Rocek:1985bk,Nishino:1991sr,Howe:1995zm} whose auxiliary fields consist of a complex scalar $Z$ and a real vector field $A_\mu$. We shall determine the most general form of the metric and the auxiliary fields under the assumption that at least one off-shell Killing spinor exists. We shall then impose the field equations of the extended topologically massive supergravity whose bosonic Lagrangian is given by~\cite{Rocek:1985bk,Nishino:1991sr}
\bea
{\cal L}_{\rm bos}  &=& R  - 2|Z|^2 + 2A_\mu A^\mu +2m( Z +{\rm h.c.}  )
\nn\w2
&&-\frac{1}{4\mu}\,\varepsilon^{\mu\nu\rho}\left( R_{\mu\nu}{}^{ab} \omega_{\rho ab}
+ \ft23 \omega_\mu^{ab} \omega_{\nu b}{}^c\omega_{\rho ca} \right)
+\frac{1}{\mu} \varepsilon^{\mu\nu\rho} F_{\mu\nu}A_\rho\ ,
\label{Lag1}
\eea
and study their supersymmetric solutions.
The total Lagrangian consists of three separately off-shell invariant pieces, one containing the Einstein-Hilbert term, another containing the cosmological term and the last one containing the Lorentz and Maxwell Chern-Simons terms. The scalar field continues to have an algebraic field equation that sets it to a constant but the field equation for the vector field takes the form $dA \sim \star A $. The Killing spinor equation implies the existence of a null or timelike Killing vector. In the former case, the field equations put the vector field to zero, for which the analysis reduces to that of off-shell $N=(1,0)$ supergravity as carried out in \cite{Gibbons:2008vi,Andringa:2009yc}. In the case of a timelike Killing vector, however, an interesting picture emerges when one seeks solutions for which the vector field is taken to be a linear combination of the dreibein, namely $A \sim A_a e^a$ where $A_a$ is a constant vector. The resulting solutions turn out to be squashed (also referred to as warped) $AdS_3$
in which the squashing parameter~$\nu$ is governed by the norm of the vector field, $A^2 = A_\mu A^\mu$, as
\bea
\nu^2 &\equiv& \frac14\,(\mu\ell)^2~=~1-\frac{A^2}{m^2}\;,
\eea
with $\ell=m^{-1}$\,.
More specifically, we find spacelike-squashed and timelike-stretched $AdS_3$, and $AdS_3$ pp-wave, for the spacelike, timelike and null norms, respectively. Similar solutions have been investigated in \cite{Anninos:2008fx} for topologically massive gravity (\ref{Lag1}) in the absence of the vector field in which case they are not supersymmetric. In that case, critical behaviour has been observed at $\mu\ell=3$ where
the solution is round $AdS_3$ and the warping transitions from stretching to squashing. Interestingly, in presence of the vector field we find that this transition occurs at $\mu\ell=2$ where the solution is null warped $AdS_3$.

Having found supersymmetric warped $AdS_3$ solutions, it is natural to examine their modding out by suitable discrete subgroups of the isometry group to obtain black hole solutions. This procedure has been systematically analyzed in \cite{Anninos:2008fx} where a black hole solution is obtained from the discrete quotient of the spacelike-stretched $AdS_3$ solution. A special case known as self-dual solution was previously found in \cite{Coussaert:1994tu}. In presence of the vector field no spacelike-stretched $AdS_3$ solution exists. Nonetheless, we find spacelike-squashed $AdS_3$ solutions which can be modded out by a suitable discrete subgroup, yielding an extremal black hole solution without closed timelike curves.

The results of this paper raise a number of interesting research topics. We shall touch on these in the final section of the paper, which is organized as follows. The model is presented in section~2, the analysis of the Killing spinor equation in section~3, the supersymmetric warped $AdS_3$ solutions in section~4, the black hole solution in section~5 and concluding remarks  in section~6. In the appendix, we present the solution of the Killing spinor equation and address the question of supersymmetry enhancement.

%%%%%%%%%%%%%%%%%%%%%%%%%%%%%%%%%%%%%%%%%%%%%%%%%%%%%%%%%%%%%%%%

\section{Massive Off-shell $(1,1)$ supergravity}

%%%%%%%%%%%%%%%%%%%%%%%%%%%%%%%%%%%%%%%%%%%%%%%%%%%%%%%%%%%%%%%%

The off-shell $(1,1)$ supergravity in $D=3$ consists of the set of fields $\{e_\mu^a, \psi_\mu, A_\mu, Z\}$ where $e_\mu^a$ is the dreibein, $\psi_\mu$ is the gravitino which is a complex Dirac spinor, $A_\mu$ is a real auxiliary vector field and $Z$ is a complex auxiliary field. The supersymmetry transformation rules, up to cubic in fermion terms, are \cite{Rocek:1985bk,Nishino:1991sr,Cecotti:2010dg}
\bea
\delta e_\mu^a &=& \ft12\bar\epsilon \gamma^a\psi_\mu + {\rm h.c.}
\nn\w2
\delta \psi_\mu &=& D_\mu\epsilon -  \ft{i}{2} A_\nu \gamma^\nu \gamma_\mu \epsilon+ \ft12 Z \gamma_\mu \epsilon^\star\ ,
\nn\w2
\delta A_\mu &=& \ft{i}{8} \bar\epsilon\gamma^{\nu\rho}\gamma_\mu \left(\psi_{\nu\rho} -iA_\lambda\gamma^\lambda\gamma_\nu \psi_\rho
+Z\gamma_\nu \psi_\rho^\star\right) + {\rm h.c.}\ ,
\nn\w2
\delta Z &=& \ft14 \overline{\epsilon^\star}\, \gamma^{\mu\nu}\left(\psi_{\mu\nu} -iA_\rho\gamma^\rho\gamma_\mu \psi_\nu
+Z\gamma_\mu \psi_\nu^\star\right)\ ,
\label{susy3}
\eea
where
\be
\psi_{\mu\nu}= 2D_{[\mu} \psi_{\nu]} \ , \qquad D_\mu \epsilon= \left(\partial_\mu +\ft14\omega_\mu{}^{ab}\gamma_{ab} \right)\epsilon\ .
\ee
We are using the conventions of \cite{Cecotti:2010dg}, and for Dirac $\gamma$-matrices we take $\gamma^\mu=(i\sigma^2, \sigma^3, \sigma^1)$
and $\epsilon^{012}=-1$.

The model we shall study is based on an action \cite{Rocek:1985bk,Nishino:1991sr}
\bea
S &=& \frac1{16\pi G}\,\int d^3x\,\sqrt{-g}\left( {\cal L}_{\rm bos} + {\cal L}_{\rm ferm}\right) \ ,
\label{model}
\eea
where the bosonic Lagrangian is given by
\bea
{\cal L}_{\rm bos}  &=& R  - 2|Z|^2 + 2A_\mu A^\mu +2m( Z +{\rm h.c.}  )
\nn\w2
&&-\frac{1}{4\mu}\,\varepsilon^{\mu\nu\rho}\left( R_{\mu\nu}{}^{ab} \omega_{\rho ab}
+ \ft23 \omega_\mu^{ab} \omega_{\nu b}{}^c\omega_{\rho ca} \right)
+\frac{1}{\mu} \varepsilon^{\mu\nu\rho} F_{\mu\nu}A_\rho\ ,
\eea
and the fermionic Lagrangian, up to quartic fermion terms, is given by
\bea
{\cal L}_{\rm ferm}  &=& - \left({\bar\psi}_\mu \gamma^{\mu\nu\rho} D_\nu\psi_\rho
+\ft12 m \overline{\psi_\mu^\star} \,\gamma^{\mu\nu}\psi_\nu  +{\rm h.c.}\right)
\nn\w2
&&
+ \frac{1}{4\mu}\,\varepsilon^{\mu\nu\rho} \left(R_{\rho\tau}-\ft14 R g_{\rho\tau}\right) \bar\psi_\mu\gamma^\tau\psi_\nu
 -\frac{1}{\mu}\,{\bar R}^\mu \gamma_\nu\gamma_\mu R^\nu\ .
\eea
The parameters $m$ and $\mu$ are real . Furthermore the Hodge dual of the gravitino curvature is defined by
\be
R^\mu  = \varepsilon^{\mu\nu\rho} (D_\nu-iA_\nu) \psi_\rho\ .
\ee
The field equations are
\bea
R_{\mu\nu} + 2 A_\mu A_\nu + 2m^2 g_{\mu\nu} + \frac1{\mu}\,C_{\mu\nu} &=& 0\ ,
\label{b1}\\[1ex]
\frac{1}{2\mu}\varepsilon^{\mu\nu\rho}\,F_{\nu\rho}  +A^\mu &=& 0\ ,
\label{b2}\\[1ex]
Z -m &=& 0 \ ,
\label{b3}\\[1ex]
R^\mu -\ft12 m \gamma^{\mu\nu}\psi_\nu ^\star + \frac{1}{2\mu}\, C^\mu &=&0\ ,
\label{f}
\eea
where\footnote{
Note that any solution of our bosonic  equations will also solve the field equations of a different model where the vector mass term in the action is replaced by Maxwell kinetic term with appropriate normalization, since the resulting Einstein equation will be the same and the vector field equation will be the derivative of ours. Such bosonic models and their solutions have been considered in \cite{Moussa:1996gm,Moussa:2008sj,Garbarz:2008qn}.}
\bea
C_{\mu\nu} &=& \varepsilon_\mu{}^{\rho\sigma}
\nabla_\rho (R_{\sigma\nu} -\ft14 g_{\sigma\nu} R)\ ,
\label{C1}\w2
C^\mu &=&  \gamma^\rho \gamma^{\mu\nu} D_\nu R_\rho
-\epsilon^{\mu\nu\rho} \left(R_{\rho\sigma} -\ft14 g_{\rho\sigma}\right) \gamma^\sigma \psi_\nu \ .
\label{C2}
\eea
These equations reduce to those for $N=(1,0)$  topologically massive supergravity \cite{Gibbons:2008vi,Andringa:2009yc} upon setting $A_\mu=0$ , $Z^\star=Z$ and $\psi_\mu^\star=\psi_\mu$.

The $N=(1,1)$ model at hand admits $AdS_3$ with radius $\ell=m^{-1}$ as a vacuum solution, for $Z=m$ and $A_\mu=0$. The fluctuations around this vacuum, with appropriate boundary conditions, arrange themselves into discrete, unitary irreducible representations of  the $AdS_3$ group $SO(2,2)$.\footnote{For  a spectrum analysis around three-dimensional Minkowski spacetime, see \cite{Andringa:2009yc}.} These irreps are denoted by $D(E_0, s)$ where $E_0$ is the lowest energy state and $S$ is the helicity of that state. For a detailed description of the method for determining the $SO(2,2)$-irreps for a given wave operator, see, for example \cite{Deger:1998nm}. Applying this method, we find the spectrum which is summarized in the following figure.

\smallskip

%%%%%%%%%%%%%%%%%%%%
\begin{displaymath}
    \xymatrix{
                                & {}\quad D(\mu\ell +1, 2) \ar@<1ex>[dl]\ar[dr] &   \\
    {\vphantom{\Big(}}  D(\mu\ell+ \frac12,\frac32) \ar[dr]   & &{}\quad D(\mu\ell+\frac32,\frac32)\ar[dl]\\
                                   & D(\mu\ell+1,1)             &  }
\end{displaymath}
%%%%%%%%%%%%%%%%%%%%%%%%%

\bigskip

We observe that if we choose $\mu\ell=1$, which is the critical value for the so-called chiral gravity as first pointed out in \cite{Li:2008dq}, then we obtain a helicity $2$ and helicity $\frac32$ singletons (also referred to as boundary excitations, as they are frozen in the bulk), and
a massive gravitino and  massive vector with representation contents $D(\frac52,\frac32)$ and $D(2,1)$, respectively.
In addition, there will be logarithmic helicity 2 and $\frac32$ modes.
The way in which the supersymmetry transformations will act on all these modes is similar to the way they do in
critical Einstein-Weyl supergravity in four dimensions~\cite{Lu:2011mw}.

%%%%%%%%%%%%%%%%%%%%%%%%%%%%%%%%%%%%%%%%%%%%%%%%%%%%%%%%%%%%%
\section{Implications of an Off-shell Killing Spinor}
\label{sec:implications}
%%%%%%%%%%%%%%%%%%%%%%%%%%%%%%%%%%%%%%%%%%%%%%%%%%%%%%%%%%%%%

Let $\e$ be a Killing spinor of (\ref{susy3}). The defining equation for the Killing spinor is
\be
D_\mu\epsilon -  \ft{i}{2} A_\nu \gamma^\nu \gamma_\mu \epsilon+ \ft12 Z \gamma_\mu \epsilon^\star =0\ .
\label{kse}
\ee
From the symmetries of the gamma-matrices one finds
\be
\eb\es=\esb\e=0.
\ee
Let us define the spinor bilinears
\bea
\eb \e &=& -\esb\es \equiv if\ ,
\nonumber\w2
\eb \gamma^\mu \e &=&\esb\gamma^\mu \es \equiv K^\mu\ ,
\nonumber\w2
\eb\gamma^\mu \es &\equiv& L^\mu \equiv S^\mu +iT^\mu\ ,
\label{norms}
\eea
where $f$ is a {\it real} function  and $K^\mu$ is a {\it real} vector.
From Fierz identities one may show that
\bea
&&K_\mu K^\mu=-f^2\ ,\quad S_\mu S^\mu=T_\mu T^\mu=f^2\ , \quad
\nonumber\\
&&K_\mu S^\mu=K_\mu T^\mu=T_\mu S^\mu=0\ ,
\eea
One also has
\bea
&&\varepsilon^{\mu\nu\rho}K_\nu L_\rho=-if L^\mu\ ,
\nonumber\w2
&&\varepsilon^{\mu\nu\rho}L_\nu L^\s_\rho=2ifK^\mu\ .
\label{cross}
\eea
Using the Killing spinor equation one finds
\be
\nabla_{(\nu} K_{\mu)}=0\ ,
\label{KV}
\ee
thus $K^\mu$ is a Killing vector which is timelike or null.
In addition to (\ref{KV}), the Killing spinor equation yields
\bea
 \partial_{[\mu} K_{\nu]} &=&
 \varepsilon_{\mu\nu\rho}\Big(
 -f A^\rho+\frac12\left(Z L^\rho+Z^\s (L^\s)^\rho\right)\Big)\;,
 \label{diff1}\\
\nabla_\mu L_\nu&=&
- i A_\nu L_\mu {-} iA_\mu L_\nu
 +i g_{\mu\nu}L_\beta A^\beta
 +\varepsilon_{\mu\nu\rho}\,K^\rho\,Z^*
 -if Z^*\,g_{\mu\nu}
 \;,
 \label{diff2}\\
 \partial_{\mu} f &=&
 - \varepsilon_{\mu\nu\rho} A^\nu K^\rho + \frac{i}2\left(Z L_\mu-Z^\s L^\s_\mu \right)
 \;.\label{diff3}
\eea
Similar structures appear in the three-dimensional
on-shell $N=8$ supergravity~\cite{Deger:2010rb}.
The implications of these equations are different depending on the norm
of the Killing vector $K^\mu$\,. We will discuss these cases separately.

%%%%%%%%%%%%%%%%%%%%%%%%%%%%%%%%%%%%%
\subsection{Null case}
%%%%%%%%%%%%%%%%%%%%%%%%%%%%%%%%%%%%%

If $f=0$, the Killing vector field $K^\mu$ is a null vector field. In this case, equations (\ref{norms}) imply that the spinor $\epsilon$
is proportional to a real spinor $\epsilon_0$ up to a complex phase
\bea
\epsilon &=& e^{-i\theta/2}\,\epsilon_0
\;,
\eea
such that in particular
\bea
L_\mu &=& e^{i\theta}\,K_\mu
\;.
\eea
Equation (\ref{diff3}) further implies that
\bea
\Im(Z e^{i\theta})\,K_\mu + \varepsilon_{\mu\nu\rho}\,A^\nu\,K^\rho &=& 0\;.
\label{ab1}
\eea
Combining (\ref{diff1}) and (\ref{diff2}), we obtain
\bea
A_\mu &=& -\frac12\,\partial_\mu\theta\;,\qquad
K^\mu A_\mu ~=~ 0\;,
\label{null_A}
\eea
for the auxiliary vector field,
and
\bea
\partial_{[\mu} K_{\nu]} &=& \varepsilon_{\mu\nu\rho}\,K^\rho\,\Re(Z e^{i\theta})
\;.
\label{ab2}
\eea

In case of a null Killing vector field, it is convenient to choose coordinates such that the metric takes the form
\be
ds^2 = dx^2 + 2P(u,x) du dv + Q(u,x) du^2\ ,
\label{null_metric}
\ee
with $K^\mu\partial_\mu = \partial_v$\,.
In these coordinates, the above equations (\ref{ab1}), (\ref{ab2}) combine into a single equation
for the auxiliary scalar field $Z$. Together with (\ref{null_A}), we find that
the Killing spinor equations in the null case imply that the auxiliary fields
are given by
\bea
A_\mu &=& -\frac12\,\partial_\mu\theta\;,\nonumber\\
Z &=& -\frac12\,e^{-i\theta}\,\partial_x\,{\rm ln} \left( P e^{i\theta} \right)
\;,
\label{null_AZ}
\eea
in terms of an unconstrained real function $\theta(u,x)$ and the metric function $P(u,x)$\,.

%%%%%%%%%%%%%%%%%%%%%%%%%%%%%%%%%%%%%%%%%%%%%
\subsection{Timelike case}
%%%%%%%%%%%%%%%%%%%%%%%%%%%%%%%%%%%%%%%%%%%%%

For $f\not=0$, the Killing vector  $K^\mu$ is timelike, $S^\mu$ and $T^\mu$ become spacelike and orthogonal to each other.
We choose a coordinate system, in which
\bea
K^\mu=\left(\begin{array}{c}1\\ 0\\0\end{array}\right)
\;,
\eea
i.e.\ the dreibein can be parametrized as
\bea
e^\mu{}_a &=& \left(
\begin{array}{ccc}
f^{-1}&-f^{-2} V_i\\
0 & e^{-\sigma}\,\delta^\alpha{}_i
\end{array}
\right)
\ , \qquad f\equiv e^\varphi\ ,
\label{vielbein}
\eea
with all components independent of the first coordinate $t$.
In this parametrisation, equation (\ref{diff3}) yields
\bea
\partial_i f &=& e^\sigma \Big(
-\epsilon_{ij}\,fA^j +\frac{i}2\,(ZL_i-Z^* L^*_i)\Big)
\ ,\qquad \partial_i\equiv \delta_i^\alpha\partial_\alpha\ .
\label{df1}
\eea
The $[i0]$ component of (\ref{diff1}) yields
\bea
\partial_i f &=& e^\sigma \epsilon_{ij} \Big(
- fA^j +\frac{1}{2}\,(ZL^j+Z^* (L^*)^j)\Big)
\;.
\label{df2}
\eea
Comparing (\ref{df1}) and (\ref{df2}), we find the relation
\bea
L_2 &=& i L_1
\;,
\label{L12}
\eea
which together with (\ref{cross}) gives
\bea
L_a &=& e^{\varphi+ic}\,\left(\begin{array}{c}
0\\1\\ i\end{array}\right)
\;,
\eea
with a time independent real function $c$.
Solving the above equation (\ref{df2}) for $A^i$ we find
\bea
A_- ~\equiv~ A^1-iA^2 &=&
  2i e^{-\sigma}\,  f^{-1} \partial_z f +\frac{1}{f}\,ZL_1
  \;,
\label{A12}
\eea
in complex coordinates $z=x+iy$.
Moreover, the $[ij]$ component of (\ref{diff2}) can be solved for $A^0$ (in the flat basis) as
\bea
A_0&=& \frac12\, \epsilon^{ij}\,\omega_{ij0}
\;.
\label{A0}
\eea

The spin connection $\omega_{abc}$, obtained for the dreibein (\ref{vielbein})
in the flat basis, has the non-vanishing components
\bea
\omega_{00i} &=& - e^{-\sigma}\,f^{-1}\partial_i f
\;,\nonumber\\
\omega_{0ij} &=& -\omega_{ij0} ~=~ f\,e^{-2\sigma}\,\partial_{[i}\left(V_{j]}e^\sigma f^{-2}\right)
\;,\nonumber\\
\omega_{ijk} &=& 2e^{-\sigma}\,\delta_{i[j}\partial_{k]}\sigma
\;.
\label{omega}
\eea

The $00$ component of (\ref{diff2}) is identically satisfied.
Also, its $0i$ and $i0$ components are identically satisfied
upon using (\ref{A0}) and (\ref{L12}). The $ij$ component of this equation gives
\bea
\nabla_i L_j &=& i(A_i L_j+A_j L_i) - i \delta_{ij} L_k A^k - f Z^* \epsilon_{ij}
\;,
\label{Lij}
\eea
with the flat derivatives $\nabla_i \equiv e_i{}^\mu\nabla_\mu$.
Using (\ref{L12}) and (\ref{omega}),
the flat derivatives $\nabla_z L_1, \nabla_{\bar z} L_1$,
take the form
\bea
\nabla_z L_1 &=& \partial_z\left(e^{-\sigma}\,L_1\right)\;,
\nonumber\\
\nabla_{\bar{z}} L_1 &=& e^{-2\sigma}\,\partial_{\bar{z}}\left(e^{\sigma}\,L_1\right)\;,
\eea
which can be used to show that the equations
(\ref{Lij}) reduce to the single complex equation
\bea
\partial_{{z}}\left(f\,e^{\sigma-ic}\right)&=&i\,f\,Z\,e^{2\sigma}
\;.
\label{holo}
\eea
To summarize, the existence of a time-like Killing spinor is equivalent
to a dreibein of the form (\ref{vielbein}) and the auxiliary fields $A_\mu$, $Z$
given by
\bea
A_0&=& \frac12\, \epsilon^{ij}\,\omega_{ij0}\;,
\label{A00}\\[.5ex]
A_1-iA_2 &=& i e^{-\sigma}\,\partial_z\left(\varphi-\sigma+ic \right)
\;,\label{Am}\\[.5ex]
Z  &=&  -ie^{-\sigma-ic}\,\partial_z\left(\varphi+\sigma-ic \right)
\;.
\label{AZ}
\eea
In particular, the last equation is equivalent to \eq{holo}. Furthermore, from the dreibein
\eq{vielbein}, it follows that
\be
ds^2= - e^{2\varphi} \left(dt+ B_i dx^i\right)^2 +e^{2\sigma} (dx^2+dy^2)\ ,
\label{metric}
\ee
where
\bea
B_i &\equiv& e^{\sigma-2\varphi} V_i\;,\qquad
\mbox{satisfying}\quad\epsilon^{ij}\partial_i B_j = -2A_0 e^{2\sigma-\varphi}\ .
\label{B}
\eea
Note that $\varphi, \sigma$ and $A_0$ are arbitrary time independent functions. These equations are the main result of this section. Let us stress again, that the analysis remains valid for any further extension of the model by matter and/or higher derivative couplings since we have
only made use of the off-shell Killing spinor equation.  For any solution to these equations it remains to verify the (model-dependent) field equations.

%

%%%%%%%%%%%%%%%%%%%%%%%%%%%%%%%%%%%
\section{A Class of Solutions}
%%%%%%%%%%%%%%%%%%%%%%%%%%%%%%%%%%%

Having found the most general form of the solution from the requirement that it admits a Killing spinor, we now examine the field equations \eq{b1}--\eq{b3} that follow from the specific off-shell $N=(1,1)$ supergravity model whose Lagrangian is given in \eq{model}.
As above, we treat separately the cases of null and timelike Killing vector.
In the null case, the vector field vanishes and all solutions are of pp-wave type as
determined in \cite{Gibbons:2008vi,Andringa:2009yc}.
In the timelike case,
and under the assumption that the vector field is constant in the flat basis, we shall find that the general supersymmetric solutions of the field equations correspond to warped $AdS$ spaces. Remarkably, the assumption of a warped $AdS$ metric together with a constant vector field alone fixes the solution up to the sign of the vector field. For one choice of the sign, the solution then is supersymmetric.

%%%%%%%%%%%%%%%%%%%%%%%%%%%%%%%%
\subsection{Null Killing vector}
%%%%%%%%%%%%%%%%%%%%%%%%%%%%%%%%

In this case given that $A_\mu=-\frac12 \partial_\mu\theta$, it follows from \eq{b2} that $A_\mu=0$. Consequently, the field equations  and the Killing spinor analysis reduce exactly to that of $N=(1,0)$ supersymmetric topologically massive supergravity whose most general supersymmetric solutions were determined \cite{Gibbons:2008vi,Andringa:2009yc}. With a slight change of coordinates, these solutions for generic value of $\mu \ell$ take the form
\bea
ds^2 = \ell^2 \left[ \frac{ dy^2 + dx^+ dx^- }{y^2} + (y^2)^{(\mu\ell +3)/2}\, h(x^-) \left(\frac{dx^-}{y^2}\right)^2 \right]\ .
\label{s1}
\eea
For $h(x^-)=\pm 1$ and $\mu\ell=-3$, this solution can be interpreted as the null warped $AdS_3$ metric.
For $(\mu \ell)^2=1$, the last term of the metric (\ref{s1}) can be removed by coordinate transformation, such that
the metric reduces to round $AdS_3$. At these values, additional solutions show up as
\bea
\mu\ell =1 &:& \qquad ds^2= \ell^2 \left[ \frac{dy^2 + dx^+ dx^-}{y^2} +  \left(\ln \frac{y}{\ell}\right) \,h(x^-) \, (dx^-)^2\right]\ ,
\label{s2}\w2
\mu\ell=-1 &:& \qquad ds^2 = \ell^2 \left[ \frac{dy^2 + dx^+ dx^-}{y^2} + \left(\ln \frac{y}{\ell}\right) \,h(x^-) \frac{(dx^-)^2 }{y^2}\right]\ ,
\label{s3}
\eea
where $h(x^-)$ is an arbitrary function of $x^-$.

%%%%%%%%%%%%%%%%%%%%%%%%%%%%%
\subsection{Timelike Killing vector}
%%%%%%%%%%%%%%%%%%%%%%%%%%%%%

In the timelike case for which $f\ne 0$, we shall seek solutions for which
\bea
Z=m\ , \qquad A_a = {\rm const}\ , \qquad A_2 = 0\ , \qquad c=0\ .
\eea
The equation $Z=m$ already follows from the $Z$ field equation (\ref{b3}), while the remaining equations constitute an ansatz. Note that the ansatz we have made for the vector field shows that it is constant in the flat basis.  From equations (\ref{Am}) and (\ref{AZ}) and their integrability conditions, we solve for $(\sigma, \varphi)$. It then follows that the non-vanishing components of the spin connection (\ref{omega}) in the flat basis are constant given by
\bea
\omega_{002}&=&m-A_1\;, \qquad \omega_{112}~=~-(m+A_1)\ ,
\nn\\
\omega_{120}&=&\omega_{201}~=~-\omega_{012}~=~A_0\ .
\label{omegaconst}
\eea
With $A_0$ constant, the last equation can be integrated to compute the off-diagonal component $V_i$ of the dreibein, but its explicit form is not needed in the following analysis. With (\ref{omegaconst}) the vector field equations \eq{b2} reduce to
\bea
(m+A_1)\,A_1-2\,A_0^2 &=& -\mu A_0\;,\nonumber\\
(m-A_1)\,A_0 &=& -\mu A_1
\;.
\label{AAm}
\eea
From these equations it follows that
\be
\left(A_0-\frac{\mu}{2}\right) A^2 =0\ .
\ee
Therefore, we have different classes of solutions
which we may distinguish as follows:

\begin{center}
   \includegraphics[width=15.5cm]{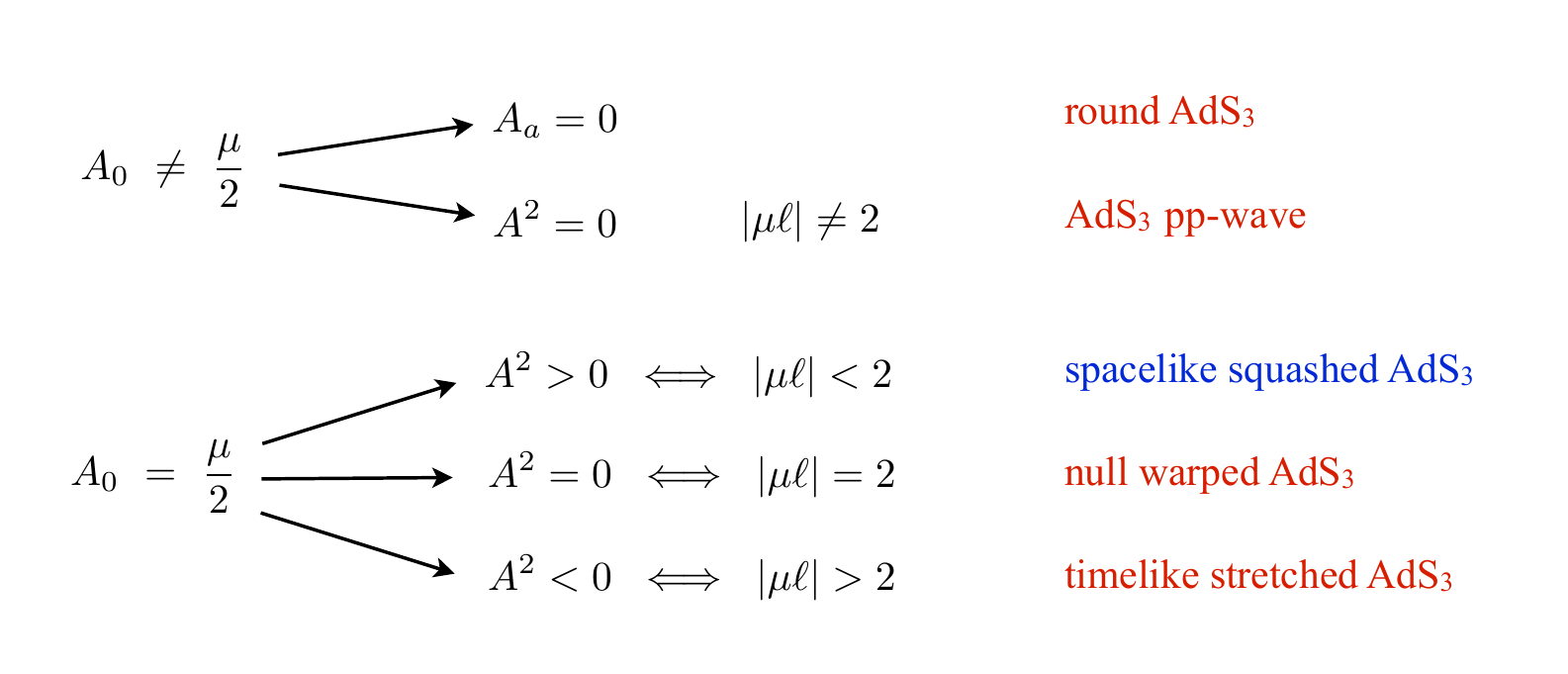}
\end{center}

\noindent
We will examine these cases separately.

%%%%%%%%%%%%%%%%%%%%%%%%%%%%%%%%%%%%%%%%%%%%%%%%%%%%%
\subsubsection{The $A_0\not=\frac{\mu}2$ case}
%%%%%%%%%%%%%%%%%%%%%%%%%%%%%%%%%%%%%%%%%%%%%%%%%%%%%
As depicted above, this case comprises two different solutions, depending on whether or not
the vector field $A_a$ vanishes identically.

%%%%%%%%%%%%%%%%%%%%%%%%%%%%%%%%%%%%%%%%%%%%%%%%%%%%%
\subsubsection*{Round $AdS_3$}
%%%%%%%%%%%%%%%%%%%%%%%%%%%%%%%%%%%%%%%%%%%%%%%%%%%%%
For $A_a=0$, the solution is  $AdS_3$ with metric
\be
ds^2= \frac{\ell^2}{y^2} \left(-d\tau^2+dx^2+dy^2\right)\ , \qquad \ell \equiv m^{-1}\ .
\label{Poincare}
\ee
For this solution the Cotton tensor $C_{ab}=0$, the Ricci tensor is given by $R_{ab}=-2m^2\eta_{ab}$ and we have the curvature invariants
\be
R =-6\,m^2 \ ,  \quad R_{\mu\nu}R^{\mu\nu} =12\,m^4 \ ,
\quad R_{\mu\nu}R^{\nu\lambda}R_{\lambda}{}^\mu =-24\,m^6\ .
\label{inv1}
\ee
%

%%%%%%%%%%%%%%%%%%%%%%%%%%%%%%%%%%%%%%%%%%%%%%%%%%%%%
\subsubsection*{The $AdS_3$ pp-wave}
%%%%%%%%%%%%%%%%%%%%%%%%%%%%%%%%%%%%%%%%%%%%%%%%%%%%%

In this case $A^2=0$ and consequently, the vector field equations (\ref{AAm}) are solved by
\bea
A_2 &=& 0\ ,\qquad A_0 = \varepsilon A_1 = \mu +\varepsilon m \ ,\quad \varepsilon = \pm 1
\;.
\label{null}
\eea
Plugging this together with $Z=m$ into the supersymmetry equations (\ref{Am}), (\ref{AZ}),
we infer that
\bea
e^\sigma  &=& z^{-1}\ , \qquad e^{\varphi} = z^\alpha\  ,
\eea
where
\be
z \equiv (m+A_1)y\ ,\qquad \alpha\equiv \frac{A_1-m}{A_1+m}\ .
\ee
For $\mu\ell=-2\varepsilon$, we find $A_0=\frac{\mu}2$, $A_1=- m$ and the solution degenerates.
This case is part of the discussion in section~\ref{subsec:A2}.
In solving  \eq{B}, without loss of generality we choose the gauge $B_y=0$ and find
\be
B_x= -\varepsilon z^{-\alpha-1}\ ,
\ee
up to an arbitrary $x$ dependent function which we can remove in the definition of the metric \eq{metric} by redefining the time coordinate $t$. Thus, we have the metric
\be
ds^2 = z^{2\alpha} \left[ -dt + 2\varepsilon z^{-\alpha-1} dx \right] dt +  \frac{1}{(A_1+m)^2}  \frac{dz^2}{z^2}\ .
\ee
It is convenient to make the coordinate transformation and relabelings
\be
z=u^{(A_1+m)/m}\  , \quad t=x^-\ , \quad x=\varepsilon x^+/2\ .
\label{zx3}
\ee
Then the metric takes the form
\bea
ds^2 &=& \ell^2 \left[ \frac{du^2+dx^+ dx^-}{u^2} - u^{2\varepsilon (\mu \ell +2\varepsilon)}  \left(\frac{dx^-}{u^2}\right)^2 \right]
\;.
\label{pp}
\eea
It remains to verify the Einstein equations (\ref{b1}). Substituting the expressions for metric
and vector field, we find that these equations are satisfied for
\bea
\varepsilon&=&1
\;.
\eea
The metric (\ref{pp}) is an $AdS$-pp wave. In these coordinates, the vector field takes the form
\bea
A &=& (\mu\ell+1)\,u^{\mu\ell}\,dx^-\;.
\eea
The Ricci tensor and the Cotton tensor in flat basis are given by
\bea
R_{ab} &=&  -2m^2\eta_{ab} + 2\mu (\mu +  m)
\left(
\begin{array}{ccc}
1& 1& 0\\[.5ex]
1&  1& 0\\[.5ex]
0&0& 0
\end{array}
\right)\ ,
\label{RC}
\\[4ex]
C_{ab} &=& -2\mu (\mu+ m) (2\mu + m)
\left(
\begin{array}{ccc}
1& 1& 0\\[.5ex]
1&  1& 0\\[.5ex]
0&0& 0
\end{array}
\right)\ .
\eea
Using these results, we find that the lowest curvature invariants remain the same as in \eq{inv1}.

Note that after the coordinate transformation (\ref{zx3}),
the limit $\mu\ell\rightarrow-2$ is well-defined and leads to the ``minus" null warped $AdS_3$ metric
\bea
ds^2 &=& \ell^2 \left[ \frac{du^2+dx^+ dx^-}{u^2} -   \left(\frac{dx^-}{u^2}\right)^2 \right]
\;.
\label{null1}
\eea

%%%%%%%%%%%%%%%%%%%%%%%%%%%%%%%%%%%%%%%%%%%%%%%%%%%%%
\subsubsection{The $A_0=\frac{\mu}2$ case}
\label{subsec:A2}
%%%%%%%%%%%%%%%%%%%%%%%%%%%%%%%%%%%%%%%%%%%%%%%%%%%%%

In this case,  we find from equations (\ref{AAm}) the solution
\bea
A_0 = \frac{\mu}{2}\ ,\quad A_1= -m
\;.
\eea
Plugging this together with $Z=m$ into the supersymmetry equations (\ref{Am}), (\ref{AZ}),
we obtain
\bea
e^\sigma  &=& 1 \ , \qquad e^{\varphi} = e^{-2my}\ .
\eea
In solving  \eq{B}, without loss of generality we choose the gauge $B_y=0$ and find
\be
B_x= \frac{\mu}{2m}\, e^{2my} \ ,
\ee
up to an arbitrary $x$ dependent function which we can remove in the definition of the metric \eq{metric} by redefining the time coordinate $t$. Thus, we have the metric
\bea
ds^2 &=& -e^{-4my} \left(dt + \frac{\mu}{2m} \,e^{2my} \, dx\right)^2 + dx^2+dy^2
\;.
\label{metricAA}
\eea
The following discussion crucially depends on the value of the parameter
\bea
\nu^2 &\equiv& \frac14\,(\mu\ell)^2~=~1-\frac{A^2}{m^2}\;,
\label{sp}
\eea
which as we shall see will take the role of the $AdS$ warping parameter of the solutions.

%%%%%%%%%%%%%%%%%%%%%%%%%%%%%%%%%%%%%%%%%%%%%%%%%%%%%%%%%%%%%
\subsubsection*{Null warped $AdS_3$}
%%%%%%%%%%%%%%%%%%%%%%%%%%%%%%%%%%%%%%%%%%%%%%%%%%%%%%%%%%%%%
For $|\mu\ell|=2$, which implies $A^2=0$, the metric (\ref{metricAA}) becomes
\bea
ds^2 &=& -e^{-4my} dt^2 \mp 2 e^{-2my}dtdx +dy^2
\;.
\eea
Upon change of coordinates
\bea
y&=&\ell \,{\rm log}\,u\;,\qquad t~=~\ell \,x^-\;,\qquad x~=~\mp\ell \,x^+
\;,
\eea
we recognize the ``minus'' null warped $AdS_3$ metric of (\ref{null1}).

%%%%%%%%%%%%%%%%%%%%%%%%%%%%%%%%%%%%%%%%%%%%%%%%%%%%%%%%%%%%%
\subsubsection*{Spacelike squashed $AdS_3$   }
%%%%%%%%%%%%%%%%%%%%%%%%%%%%%%%%%%%%%%%%%%%%%%%%%%%%%%%%%%%%%

For $|\mu\ell| < 2$, which implies $A^2>0$, we rewrite the metric (\ref{metricAA})
as
\bea
ds^2
&=& \frac{A^2}{m^2}\left(dx-\frac{\mu m}{2A^2} e^{-2my} dt\right)^2 -\frac{m^2}{A^2} e^{-4my} dt^2 +dy^2
\;.
\eea
It is then
convenient to make the coordinate transformation
\be
t'=2mt\ ,\qquad x'=-4m \sqrt{A^2/\mu^2}\,x\ ,\qquad z=\sqrt{A^2/m^2}\,e^{2my} \ ,
\ee
so that the metric becomes
\be
ds^2 = \frac{\ell^2}{4} \left[ \frac{-dt'^2+dz^2}{z^2}+ \nu^2 \left(dx'+ \frac{dt'}{z}\right)^2 \right]\ .
\label{ss1}
\ee
This is spacelike squashed $AdS_3$ with squashing parameter $\nu^2$ given by (\ref{sp}) above.
The terminology of ``squashed'' is due to the relation $\nu^2 <1$.
The above metric can be expressed in global coordinates by the following coordinate transformation \cite{Anninos:2008fx}
\bea
t' &=& -\ft12  (\tanh \sigma -\cos \tau)^{-1}\sin \tau\ ,
\nn\w2
x' &=& u+2\tanh^{-1} (e^\sigma\,\tan \frac{\tau}{2})\ ,
\nn\w2
z &=& \ft12 (\tanh \sigma -\cos \tau)^{-1}(\cosh \sigma)^{-1} \ ,
\eea
with $u,\sigma \in \mathbb{R}$ and $\tau \sim \tau+4\pi$ \cite{Jugeau:2010nq}. The metric \eq{ss1} then takes the form
\bea
ds^2 =\frac{\ell^2}{4} \left[ -\cosh^2\sigma\,d\tau^2 +d\sigma^2 + \nu^2\,(du + \sinh\sigma\, d\tau)^2\right]\ .
\label{st2}
\eea
In these coordinates, the vector field $A_\mu$ reads
\bea
A &=& \frac12\,\nu\,\sqrt{1-\nu^2}\left({\rm sinh}\,\sigma\,d\tau+du\right)\ .
\label{vf1}
\eea
In terms of the left-invariant 1-forms $\theta^a$ obeying $d\theta_a = -\frac12 \epsilon_{abc}\, \theta^b \wedge \theta^c$, and given by
\bea
\theta^0 &=& -d\tau\, \cosh u \cosh \sigma + d\sigma\, \sinh u
\nn\w2
\theta^1 &=& d\tau\, \sinh u \cosh \sigma - d\sigma\, \cosh u
\nn\w2
\theta^2&=& d\tau\, \sinh \sigma +du\ ,
\eea
where $u,\sigma \in \mathbb{R}, \tau \sim \tau + 4\pi$, the metric (\ref{metricAA}) and \eq{vf1} can be written as
\bea
ds^2  &=& \frac{\ell^2}{4}  ( -\theta^0 \otimes \theta^0 + \theta^1 \otimes \theta^1 + \nu^2 \,\theta^2 \otimes \theta^2)\ ,
\nn\w2
A &=& \frac12\,\nu\,\sqrt{1-\nu^2}\, \theta^2\ .
\eea
Thus, it is readily seen that the solution is invariant under $SL(2,R)_R \times \mathbb{R}$, where $\mathbb{R}$ is generated by translations along $u$.
If we compactify along the $u$ direction by identifying $u \sim u+2\pi \beta$, we obtain the so-called self-dual solution \cite{Coussaert:1994tu} with
\be
ds^2 =\frac{\ell^2}{4} \left[ -\cosh^2\sigma\,d\tau^2 +d\sigma^2 + \nu^2\,(\beta\,d\phi + \sinh\sigma\, d\tau)^2\right]\ .
\label{sd}
\ee
with $\tau,\sigma \in \mathbb{R}$ and $\phi \sim \phi+2\pi$. The isometry group of this solution becomes $SL(2,R)_R\times U(1)_L$.

%%%%%%%%%%%%%%%%%%%%%%%%%%%%%%%%%%%%%%%%%%%%%%%%%%%%%%%%%%%%
\subsubsection*{Timelike stretched $AdS_3$  }
%%%%%%%%%%%%%%%%%%%%%%%%%%%%%%%%%%%%%%%%%%%%%%%%%%%%%%%%%%%%

For $|\mu\ell| > 2$, which implies $A^2<0$, starting from the metric (\ref{metricAA}), we make the coordinate transformation
\be
t'=2mt\ ,\qquad x'=-4m \sqrt{-A^2/\mu^2}\,x\ ,\qquad z=\sqrt{-A^2/m^2}\,e^{2my} \ .
\ee
Then the metric takes the form
\be
ds^2 = \frac{\ell^2}{4} \left[ - \nu^2 \left(dx'+ \frac{dt'}{z}\right)^2+\frac{dt'^2+dz^2}{z^2} \right]\ .
\ee
This is timelike stretched $AdS_3$ with squashing parameter $\nu^2$, again given by \eq{sp}.
The terminology of ``stretched'' is due to $\nu^2 >1$.
The above metric can be expressed in global coordinates by the following coordinate transformation
\bea
t' &=& -\ft12  (\tanh \sigma -\cosh u)^{-1}\sinh u
\nn\w2
x' &=& \tau+2\tan^{-1} (e^\sigma\,\tanh \frac{u}{2})
\nn\w2
z &=& \ft12 (\tanh \sigma -\cosh u)^{-1}(\cosh \sigma)^{-1}
\eea
In these coordinates, the  timelike stretched $AdS_3$ solution takes the form
\be
ds^2 = \frac{\ell^2}{4} \left[ -\nu^2(d\tau+ \sinh \sigma\,du)^2 +d\sigma^2 + \cosh^2 \sigma du^2\right]\ .
\label{metricBB}
\ee
In these coordinates, the vector field $A_\mu$ reads
\bea
A &=& \frac12\,\nu\,\sqrt{\nu^2-1}\left(d\tau+{\rm sinh}\,\sigma\,du\right)\ .
\label{vf2}
\eea
In terms of the right-invariant 1-forms ${\tilde\theta}^a$ obeying
$d{\tilde\theta}_a = +\frac12 \epsilon_{abc}\, {\tilde\theta}^b \wedge {\tilde\theta}^c$, and given by
\bea
{\tilde\theta}_0 &=& d\tau +du\, \sinh \sigma\ ,
\nn\w2
{\tilde\theta}_1 &=& du\,\cos\tau \cosh \sigma + d\sigma\, \sin \tau
\nn\w2
{\tilde\theta}_2&=& du\, \sin \tau  \cosh \sigma - d\sigma\, \cos \tau\ ,
\eea
the metric (\ref{metricBB}) and vector field \eq{vf2} can be written as
\bea
ds^2 &=& \frac{\ell^2}{4}  ( - \nu^2\, {\tilde\theta}_0 \otimes {\tilde\theta}_0
+ {\tilde\theta}_1 \otimes {\tilde\theta}_1 + {\tilde\theta}_2 \otimes {\tilde\theta}_2)\ ,
\nn\w2
A &=& \frac12\,\nu\,\sqrt{\nu^2-1}\,{\tilde\theta}_0\ .
\eea
Thus, we see that the solution is invariant under $SL(2,R)_L \times U(1)_R$.
\smallskip

For both, the spacelike  squashed and timelike stretched $AdS_3$, solutions given above, the Ricci tensor and the Cotton tensor in the flat basis are given by
\bea
R_{ab} &=&  -2m^2\,(2-\nu^2) \,\eta_{ab} + 4m^2
\left(
\begin{array}{ccc}
 {  \nu^2}& -\nu & 0\\[.5ex]
-\nu & 1& 0\\[.5ex]
0&0& 0
\end{array}
\right)\ ,
\label{RC2}
\\[4ex]
C_{ab} &=& 4\,m^3\,\nu\,(1-\nu^2) \,\eta_{ab} -12\, m^3\,\nu \left(
\begin{array}{ccc}
\nu^2& -  \nu& 0\\[.5ex]
- \nu &  1& 0\\[.5ex]
0&0& 0
\end{array}
\right)\ .
\eea
It follows that the lowest curvature invariants in these cases differ from the ones (\ref{inv1}) for  $AdS_3$ and they are given by
\bea
&& R =-2m^2 (4-\nu^2)\ , \quad R_{\mu\nu}R^{\mu\nu} =4m^2( 8-8\nu^2+3\nu^4)\ ,
\nonumber\w2
&&{}
R_{\mu\nu}R^{\nu\lambda}R_{\lambda}{}^\mu = 8m^6 ({ -} 16+24\nu^2-12\nu^4+\nu^6)\  .
\eea

%%%%%%%%%%%%%%%%%%%%%%%%%%%%%%%%%%%%%

\section{Extremal Black Hole}

%%%%%%%%%%%%%%%%%%%%%%%%%%%%%%%%%%%%%

The self-dual solution given in \eq{sd} can be interpreted as an extremal black hole with horizon radius $r_h$.  It is represented conveniently in the Schwarzschild coordinates $(t,r,\theta)$ as follows
\bea
ds^2
&=&
\frac{3\ell^2}{4-\nu^2} \left[ \Big(dt + \frac{\sqrt{3}(\nu r -r_h)}{\sqrt{4-\nu^2}} \,d\theta\Big)^2
+\frac{4-\nu^2 }{12\,(r-r_h)^2}  \, dr^2
-\frac{3(r-r_h)^2}{4-\nu^2}\,d\theta^2\right]\ ,
\nonumber\\
\label{ct4}
\eea
and the vector field is given by
\bea
A &=& \sqrt{\frac{3\,(1-\nu^2)}{4-\nu^2}}\,dt+3\,(\nu r+r_h)\frac{\sqrt{1-\nu^2}}{4-\nu^2}\,d\theta\;.
\eea
The metric (\ref{ct4}) has no causal or geometric singularities. However, it has a Killing horizon at $r=r_h$ associated with the 
Killing vector of the time translations,  where $g^{rr}$ and the determinant of $(t,\theta$) part of the metric vanishes. The coordinate transformation that maps this metric to the one in the global coordinate system in
(\ref{st2}) is given by
\bea
t &=& -\nu\, \sqrt{\frac{(4-\nu^2)}{12}}  \left[u+2\tanh^{-1}\left(e^\sigma\,\tan \frac{\tau}{2} \right)\right]
-\frac{r_h \sqrt{4-\nu^2}\,(1+\nu)\,\sin\tau}{4\sqrt{3}\, (\tanh\,\sigma -\cos\,\tau)}\ ,
\nn\w2
\theta &=& -\frac{(4-\nu^2) \sin\,\tau}{12\,(\tanh\,\sigma - \cos\,\tau)}\ ,
\nn\w2
r&=& 2\,(\sinh\,\sigma -\cos\,\tau\,\cosh\,\sigma) + r_h \ .
\label{ct5}
\eea
The formulae \eq{ct4} and \eq{ct5} are obtained from those given in \cite{Anninos:2008fx} by suitable adaptation to our spacelike squashed $AdS_3$ metric, taking into account the differences in the radius and squashing parameter. 
In (\ref{ct4}) the coordinate $r>0$ 
and $\theta \in \mathbb{R}$.~\footnote{Since $\nu^2<1$ for the self-dual solution,
$\theta$ cannot be identified periodically, which would imply the existence of closed time-like curves for large $r$. }
Recall that the $u$ coordinate is compactified by the relation
$u\simeq u+2\pi \beta$, thus $t\simeq t-\nu\, \sqrt{\frac{(4-\nu^2)}{3}} \pi \beta$. 
The corresponding $SL(2,R)_L$ isometry is generated by the Killing vector $\partial_u$.
Following~\cite{Anninos:2008fx}, we observe that the right temperature $T_R$ is vanishing
while the left temperature is read off from
\bea
\frac{\partial}{\partial t} &=&  \frac{\pi\ell}{\beta}\,T_L\,\frac{\partial}{\partial u}
\;.
\eea
From \eq{ct5}, this leads to
\bea
T_L &=& \frac{2\sqrt{3}\,\beta}{\sqrt{4-\nu^2}\,\nu\pi\ell}
\;.
\eea
To compute the entropy we follow the method of~\cite{Sahoo:2006vz}.
Expressing the metric as
\bea
ds^2 &=& G_{MN}\, dx^M dx^N~=~\phi\left( g_{\mu\nu}\,dx^\mu dx^\nu+(dt+A_\mu dx^\mu)^2\right)
\;,
\eea
the contribution from the gravitational Chern-Simons term is given by
\bea
 S_{\rm LCS}&=& \frac{\pi}{8\mu G}\,\frac{\epsilon^{\mu\nu}F_{\mu\nu}}{\sqrt{-g}}
 \;,
 \eea
while the contribution from the remainder of the action is given by
\bea
S_{\rm R}&=& \frac1{4G}\,\int dt \sqrt{G_{tt}}
\;.
\eea
This leads to
\bea
S &=& S_{\rm R} +S_{\rm LCS}~=~ \frac{\pi\beta\nu\ell}{4G}+\frac{\pi\beta\nu\ell}{8G}
\;.
\eea
From a dual CFT consideration, assuming that the total entropy is given by
\bea
S&=&\frac{\pi^2\ell}{3}\,T_L\,c_L
\;,
\eea
we deduce that
\bea
c_L &=&
\frac{3\nu^2\ell}{16\,G} \,
\sqrt{3\,(4-\nu^2)}
\;.
\eea

%%%%%%%%%%%%%%%%%%%%%%%%%%%%%%
\section{Comments}
%%%%%%%%%%%%%%%%%%%%%%%%%%%%%%

In this paper we have constructed supersymmetric warped $AdS_3$ solutions in
off-shell extended topologically massive supergravity.
The auxiliary vector field of $N=(1,1)$ supergravity is a topologically massive vector
and plays a crucial role for the existence of these solutions. In particular, its norm is related
to the warping parameter of the solutions. At vanishing norm, which corresponds to $\mu\ell=2$,
the solution is null warped $AdS_3$
which represents the transition point between spacelike squashed and timelike stretched solutions.
In contrast, in absence of the vector field, non-supersymmetric warped $AdS$ solutions exist \cite{Anninos:2008fx} whose critical point occurs at $\mu\ell=3$ where the solution is round $AdS_3$. In \cite{Nilsson:2013fya} it is shown that only for certain values of $\mu\ell$, which are obtained by varying the number of scalar fields with a VEV,
solutions of topological gauged $D=3$, $\mathcal{N}=8$  CFTs
with SO(N) gauge group are also solutions of topologically massive gravity. It is interesting to note that among these, $\mu\ell=2$ is the only even one and until now no solution with this value were known. Supersymmetry of these warped $AdS_3$ solutions may also prove useful to further analyze the
stability of them, c.f.~\cite{Anninos:2009zi}.
Furthermore, we have constructed an extremal black hole solution as a discrete quotient of
spacelike squashed warped $AdS_3$. This solution has no closed timelike curves.
It would also be interesting to study its holographic dual as in \cite{Anninos:2008qb}.

Among our key results are the expressions for the most general form of the metric and auxiliary fields that follow from the Killing spinor equation. To obtain these results, it was sufficient to study the direct consequences of the Killing spinor equation itself, as summarized in equations \eq{KV}--\eq{diff3}.
Nonetheless, it is useful to record the integrability conditions that follow from the Killing spinor equations. To this end, we note that acting with $\bar\epsilon\gamma^{\mu\nu} D_\nu$ on the Killing spinor equation (\ref{kse}) implies\footnote{ A constraint on the curvature scalar has been noted in off-shell $N=1, D=10$ supergravity in a different context, resulting from its superspace formulation \cite{Nilsson:1985si,Howe:1986ed}.}
\bea
R+2 A_\mu A^\mu+6 |Z|^2 -8\,\Im\left(e^{-\sigma+i c}\, \partial_{\bar{z}} Z\right) &=& 0
\;,
\nonumber\\[.5ex]
\nabla_\mu A^\mu -2 \Re\left(e^{-\sigma+i c}\, \partial_{\bar{z}} Z\right) &=& 0\ .
\label{int}
\eea
It is interesting that universal equations involving the Ricci scalar and the divergence of the vector field arise, regardless of the Lagrangian field theory one may consider, c.f.~(\ref{int}).\footnote{By contrast, Killing spinor equations resulting from on-shell supergravity are known to imply a subset of the field equations.} To determine the consequences of these equations when considered together with the field equations in general, let us consider the extension of the model we have studied here by adding to it a new piece  ${\cal L'}$ that may contain all possible higher derivative terms allowed by off-shell supersymmetry. Focusing on the bosonic sector, the resulting field equations together with the integrability conditions above give
\bea
&& g^{\mu\nu} \frac{\delta\cal L'}{\delta g^{\mu\nu}} + \frac32 \left( Z \frac{\delta\cal L'}{\delta Z} + h.c.\right) = 4 \Im\left(e^{-\sigma+i c}\, \partial_{\bar{z}} Z\right)\ ,
\nonumber\\[.5ex]
&& \nabla^\mu \frac{\delta\cal L'}{\delta A_\mu} = -8 \Re\left(e^{-\sigma+i c}\, \partial_{\bar{z}} Z\right)\ .
\eea
It may be worthwhile to investigate how powerful these conditions are on $\cal L'$. It would also be interesting to explore the relation between the off-shell theories studied here and and the on-shell ones that arise in spontaneous compactification of string theory that preserves as much supersymmetry.

The analysis which we have done in this paper in fact has direct applications for the construction of rigid
supersymmetric theories on curved spacetimes. Along the lines of \cite{Festuccia:2011ws} every solution to the off-shell Killing spinor equations defines a background on which a rigid supersymmetric field theory can be constructed. It suffices to evaluate any off-shell matter coupling to the Lagrangian (\ref{model}) on the background metric (\ref{metric}) with auxiliary fields (\ref{AZ}).
A similar analysis (based on new minimal off-shell supergravity in $D=3$)
has been done for Euclidean backgrounds in \cite{Closset:2012ru},
and for superconformal theories in \cite{Klare:2012gn,Hristov:2013spa}.
See  \cite{Kuzenko:2011rd,Kuzenko:2011xg} for the corresponding constructions in superspace.

An interesting open problem is the possible embedding of our solutions into string theory.
It is known that a suitable quotient of spacelike squashed $AdS_3$, known as G\"odel black hole, complemented by a suitable internal space, arises as a supersymmetric solution of Type II supergravities truncated to the Neveu-Schwarz sector \cite{Compere:2008cw}. It is not clear what three-dimensional supergravity it solves, and it harbors closed timelike curves in the asymptotic region.
Finally we comment on a consistent truncation of Type IIB supergravity that has been shown to yield a Lagrangian in three dimensions whose bosonic sector consist of graviton, two real scalars and a real vector field \cite{Detournay:2012dz}. This is not a bosonic sector of any supergravity theory in three dimensions (but fits into $N=2$ supergravity upon retaining more fields in the reduction~\cite{Karndumri:2013dca}), and it does not contain a Lorentz-Chern-Simons term. Nonetheless it is intriguing that the Lagrangian for the vector fields is schematically given by $F\wedge A + A^2$ (with scalars taken to be constant), just as in the model we have studied here. Warped black string solution and a  holographic description of the theory have been discussed in \cite{Guica:2013jza}. To examine the relation of this model (or more generally the suitably compactified on-shell ten-dimensional supergravities) with the off-shell model we have studied here, it would be useful to extend it by constructing all possible
curvature square invariants as a start.

\subsection*{Acknowledgments}

We would like to thank David Chow, Geoffrey Comp\`ere, Monica Guica, \, Murat G\"unaydin, Wei Li, Yi Pang, Chris Pope, Wei Song and Per Sundell for useful discussions. The research of N.S.D. is partially supported by TUBITAK grant 113F034. The research of A.K. is partially supported by TUBITAK-BIDEB 2219 grant.
The research of E.~S. is supported in part by NSF grant PHY-0906222 and PHY-1214344.

%\bigskip
%\bigskip

\begin{appendix}

%%%%%%%%%%%%%%%%%%%%%%%%%%%%%%%%%%%%%%%%%%%%%%%%%%%%%%%%%%%
\section{Solution of the Killing spinor equation}
%%%%%%%%%%%%%%%%%%%%%%%%%%%%%%%%%%%%%%%%%%%%%%%%%%%%%%%%%%%

In this appendix, we integrate the Killing spinor equations and explicitly determine the Killing spinor
in the background found in section~\ref{sec:implications}.
The Killing spinor equations translate into
\bea
0 &=&
e^{-\sigma}\,\gamma^0\,\partial_t\,\epsilon + A_0\,\bar{\mathbb{P}}\,\epsilon
+\frac{i}2\left(4A_{\bar{z}}+ e^{-ic}Z^*\right) \bar{\Omega}\,\epsilon
+\frac{i}2\,e^{ic}Z\,\Omega\,\epsilon+\frac12\,Z\,\epsilon^*
\;,\nonumber\\
0 &=&
e^{-\sigma}\Big(\partial_z+\frac{i}2\,\partial_z\sigma\,\gamma^0\Big)\epsilon
-A_0\,\bar{\Omega}\,\epsilon-i\,A_z\,\bar{\mathbb{P}}\,\epsilon+\frac12\,Z\,\bar{\Omega}\,\epsilon^*
\;,\nonumber\\
0 &=&e^{-\sigma}\Big(\partial_{\bar{z}}-\frac{i}2\,\partial_{\bar{z}}\sigma\,\gamma^0\Big)\epsilon
-i\,A_{\bar{z}}\,{\mathbb{P}}\,\epsilon+\frac12\,Z\,\Omega\,\epsilon^*
\;,
\label{kse_exp}
\eea
where we have defined the combinations of $\gamma$ matrices
\bea
\mathbb{P} \equiv  \frac12\left(\,\mathbb{I}_2 + i\,\gamma^0\right)\;,\qquad
\Omega \equiv \frac12\left(\gamma^1+i\gamma^2\right) = \gamma^1 \mathbb{P}
\;.
\eea
An immediate solution to (\ref{kse_exp}) is provided by
\bea
\epsilon&\equiv& f^{1/2}\,e^{-ic/2}\,\epsilon_0
\;,
\label{eps}
\eea
with a constant spinor $\epsilon_0$ satisfying the conditions
\bea
\bar{\mathbb{P}}\,\epsilon_0 &=& 0\ ,\qquad  \epsilon_0^* = -i\,\Omega\,\epsilon_0\ .
\label{pc1}
\eea
With (\ref{eps}), we have thus reconstructed the original Killing spinor from
which the spinor bilinears~(\ref{norms}) were computed.

Next, let us examine whether the Killing spinor equations (\ref{kse_exp}) may
admit additional solutions. For round $AdS_3$, in which case $Z=m$ and $A_a=0$, it is well known that the
Killing spinor equations \eq{kse} admit a second solution of the type
\be
\mathbb{P}\,\epsilon=0\ .
\label{cc2}
\ee
In the general case, for non-vanishing $A_0$, one can show that equations (\ref{kse_exp}) do not admit additional
solution of this type. Namely, for (\ref{cc2}) the second equation in \eq{kse_exp} takes the form
\be
{\cal O} \epsilon= A_0 {\bar\Omega}\epsilon\ ,
\label{c1}
\ee
where ${\cal O}$ is a differential operator which is proportional to unity in spinor space and whose exact form can be read of from \eq{kse_exp}. Multiplying this equation with $\mathbb{P}$ from the left, and using \eq{cc2} and $\mathbb{P}{\bar\Omega}={\bar\Omega}$, we find that $A_0 {\bar\Omega}\epsilon=0$. Multiplying this equation by $\Omega$ from the left and noting that  $\Omega {\bar\Omega} = \bar{\mathbb{P}}$,  we find $\bar{\mathbb{P}}\epsilon=0$, assuming that $A_0$ is nonvanishing. Together with \eq{cc2}, it follows that $\epsilon=0$.

Relaxing (\ref{cc2}), leads to coupling between $\mathbb{P}\,\epsilon$ and $\bar{\mathbb{P}}\,\epsilon$,
but we still find that for all our solutions other than $AdS_3$ these equations do not admit a solution. Thus there is
no supersymmetry enhancement except for the $AdS_3$ case.

\end{appendix}

%\bigskip
%\bigskip

%%%%%%%%%%%%%%%%%%%%%%%%%%%%%%%%%%
%
%\bibliographystyle{JHEP2}
%\bibliography{refs}
%\end{document}
%
%%%%%%%%%%%%%%%%%%%%%%%%%%%%%%%%%%
\providecommand{\href}[2]{#2}\begingroup\raggedright\endgroup

\end{document}